\documentclass[fleqn,twoside]{article}
\usepackage{amssymb}
\usepackage{graphicx}
\usepackage{espcrc2}
\title{
\vspace{-0.5cm}
\begin{flushright}
{\normalsize CPT-2004/P.052}\\
{\normalsize SHEP-0428}
\end{flushright}
\vspace{-0.5cm}
Lattice artefacts in SU(3) lattice gauge theory with a mixed
  fundamental and adjoint plaquette action}
\author{Martin Hasenbusch\address{Department of Physics and Astronomy,
  University of Southampton, Southampton SO17 1BJ, UK}, Silvia
Necco\address{Centre de Physique Th\'eorique, CNRS Luminy, F-13288 Marseille
  -- Cedex 9, France}\thanks{Supported by TMR,
EC-Contract No. HPRNCT-2002-00311 (EURIDICE)}}
\begin{document}
\vspace{-0.8cm}
\begin{abstract}
We investigated SU(3) lattice gauge theory with a fundamental and adjoint
 plaquette term in the action. The purpose is to test whether the choice of a
negative adjoint coupling can reduce lattice artefacts and improve the scaling behaviour. To this end, we  have studied
the finite temperature phase transition, the static
potential and the mass of the $0^{++}$ glueball.
We found that indeed the lattice artefacts
in e.g. $m_{0^{++}}/T_c$ can be reduced considerably compared with the
pure Wilson (fundamental) gauge action at the same lattice spacing.
\vspace{-0.5cm}
\end{abstract} 
\vspace{-2cm}
\maketitle
\vspace{-3cm}
\section{INTRODUCTION}
The choice of the gauge SU(3) action plays an important role in QCD
lattice simulations, and many efforts have been spent in searching for a
formulation where discretisation errors are reduced and/or topological
dislocations are suppressed. 
In this work we investigate a gauge action containing plaquettes in both
fundamental and adjoint representation. The motivation for this choice is that
in the $(\beta_f,\beta_a)$ plane, the pure SU(3) gauge theory in four
dimensions has a line of first order phase transitions with an endpoint at 
   \cite{Blum:1995xb}
\begin{equation}
\label{endpoint}
 (\beta_f,\beta_a) = (4.00(7),2.06(8)) \;\;\;.
\end{equation}
The presence of this transition line causes large lattice artefacts;
in particular, the mass 
$m_{0^{++}}$ of the lightest glueball $0^{++}$  is expected to go to zero as the end-point is approached \cite{Heller}. As a relic of this 
behaviour, 
the estimate of $m_{0^{++}}$ in some physical unit 
at $\beta_a=0$, for $5.5 < \beta_f < 6.0$ is much 
smaller than the continuum result.
The purpose of this work is to study whether the discretisation errors can be
reduced by choosing a negative value of $\beta_a$, i.e. by moving away from
the transition line.
\vspace{-1cm}
\section{THE ACTION}
We consider lattice SU(N) gauge theory with a mixed fundamental-adjoint action,
\begin{equation}\label{mixed_action}
S \;=\; \beta_f \; \sum_P \left[1-\frac{1}{N} \mbox{Re} \mbox{Tr}_f U_P \right]
      \;+
\end{equation}
$$
\; \beta_a \; \sum_P
 \left[1-\frac{1}{N^2} \mbox{Tr}_f U_P^{\dag} \mbox{Tr}_f U_P \right]\;\;,
$$
where $\beta_f$, $\beta_a$ are the fundamental and adjoint couplings and
$U_P$ is the elementary plaquette. We adopted 4-dimensional lattices with spatial extensions
$aN_s$ and temporal extension $aN_t$, with periodic boundary conditions in all
directions. The Yang-Mills theory is recovered in the naive continuum limit 
if the bare coupling $g_0$ is defined by $6/g_0 ^2=\beta_f+2\beta_a$.

Details on the simulation
algorithm and its performance are given in \cite{martin_silvia}.
\vspace{-0.1cm}
\section{THE FINITE TEMPERATURE PHASE TRANSITION}
For a lattice with $N_t$ points in the time direction, in the 
limit $N_s \rightarrow \infty$,
the deconfinement temperature 
is given by
\begin{equation}
\frac{1}{T_c}=N_t a(\{\beta_f,\beta_a\}_c) \;\;,
\end{equation}
where $\{\beta_f,\beta_a\}_c$ indicates the critical coupling.
In our study we kept fixed the adjoint coupling to $\beta_a=0, -2.0, -4.0$ and
determined $\beta_{f,c}$ for $N_t=2,3,4,6$ by adopting the method discussed in \cite{Borgs}.

Our results for $\beta_{f,c}$ are summarised in Table \ref{ourTc};
for $\beta_a=0$ we find a substantial agreement with other results present in
the literature \cite{AlBeSa92,QCDPAX,Bielefeld96,Beinlich:1997ia}.
 For $\beta_a=0$, $N_t =6$ we performed no own simulation, but used in the
 following the result $\beta_{f,c} = 5.89405(51)$ of ref. \cite{QCDPAX}.
\begin{table}
\caption{\sl \label{ourTc}
Numerical results for the finite temperature phase transition obtained for
the fundamental-adjoint gauge action.}
\begin{center}
\begin{tabular}{|l||l|l|l|}
\hline
 $N_t$;  $\beta_a$ &\phantom{00} 0.0&\phantom{0}--2.0 &\phantom{0}--4.0 \\
\hline
\hline
\phantom{0}  2   & 5.0948(6) &  6.4475(6)  &  7.8477(6) \\
\hline
\phantom{0}  3   & 5.5420(3) &  7.1603(3)  &  8.8357(4) \\
\hline
\phantom{0}  4   & 5.6926(2) &  7.4433(3)  &  9.2552(6) \\
\hline
\phantom{0}  6   & \phantom{000}-  & 7.8056(5) & 9.7748(11) \\
\hline
\end{tabular}
\end{center}
\vspace{-1cm}
\end{table}
\vspace{-0.2cm}
\section{THE STATIC POTENTIAL} 
In order to fix the scale we computed the static potential (at zero
temperature) and extracted the string tension $a^2\sigma$.
The static potential has been computed through the Polyakov loop correlation function:
\begin{equation}
a V(r)=-\frac{1}{N_t} \left[ \log\langle P(x)^* P(y)\rangle +\epsilon \right]
\;\;,
\end{equation}
where $y=x+r\hat{1}$ and
$\epsilon$ is the correction due to excited states in the string spectrum.
Here we used a large temporal extension $a N_t >> 1/T_c$ to ensure that
finite-$N_t$ corrections are negligible; in particular we adopted $N_t = 6/(a
T_c)$ for all computations.
The string tension has then been evaluated from the ansatz
\begin{equation}
\label{pot}
V(r)=\sigma r + \mu - \frac{\pi}{12 r}\left(1+\frac{b}{r}\right),
\end{equation}
where $b$ has been recently shown from theoretical principles to be 
zero \cite{lw_new}.\footnote{In our evaluation we considered $b=0.04{\rm fm}$,
which was evaluated numerically in \cite{lw}. Nevertheless the effect of
having $b\neq 0$ is only minor and not crucial for our computation of
$\sigma$.}  Moreover, we adopted the tree-level improved distance $r_I$,
defined such that the force at tree-level has no lattice artefacts.
In order to compute the static potential up to large distances by keeping the
statistical uncertainties under control, we adopted a variant
of the algorithm proposed by L\"uscher and Weisz \cite{lw}.
The main difference consists in using  factorisation in the spatial
directions, in addition to the one in the temporal direction.
The full details of the procedure are given in \cite{martin_silvia}. 
For $1/(aT_c)=4$ we were able to extract $a^2\sigma$ from the force at
$r/a=7$ ($\beta_a=0$); in the other cases our final values were taken from
$r/a=4$ and $r/a=6$. In all computations we where confident the quoted error also
covers possible systematic uncertainties.
The numerical results for $a^2\sigma$ are reported in
\cite{martin_silvia}.

\begin{figure}
\includegraphics[width=7.5cm]{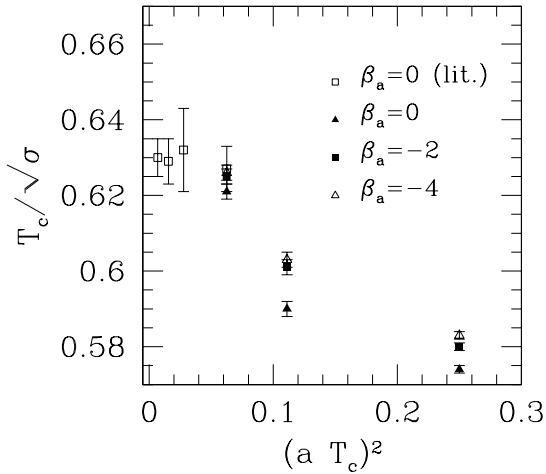}
\vspace{-1.5cm}
\caption{\sl Results for $T_c/\sqrt{\sigma}$ as a function 
of $(aT_c)^2$. In addition, we report results  from  
\protect\cite{Beinlich:1997ia} for the  Wilson action at smaller lattice spacings.}\label{fig_string}
\vspace{-0.5cm}
\end{figure}

The dimensionless quantity $T_{c}/\sqrt{\sigma}$ as function of the lattice
spacing is  plotted in fig. \ref{fig_string},
 together with other values obtained for the Wilson 
action \cite{Beinlich:1997ia}. 
We notice that for $\beta_a=-2$ and $-4$ the estimate for $T_c/\sqrt{\sigma}$
is closer to the continuum limit than for $\beta_a=0$.  However,
the difference between $1/(a T_c)=3$ and $1/(a T_c)=4$ is larger than
that for the different values of $\beta_a$ at fixed $1/(a T_c)$.
\vspace{-0.1cm}
\section{THE $0^{++}$ GLUEBALL MASS}
\begin{figure}
\includegraphics[width=7.5cm]{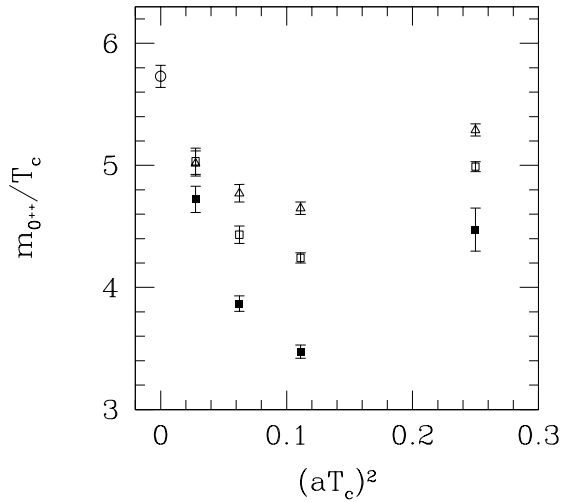}
\vspace{-1.2cm}
\caption{\sl $m_{0^{++}}/T_c$ for $\beta_a=0$ (filled squares) 
$\beta_a=-2$ (open squares) and $\beta_a=-4$ (triangles)
as function of $(aT_c)^2$. The circle 
gives the continuum results extracted from the literature.}\label{mass_scaling}
\vspace{-0.5cm}
\end{figure}

The mass of the lightest glueball is expected to be most sensitive
to the choice of the action in the $(\beta_f,\beta_a)$ plane.
The $0^{++}$ glueball mass has been computed through the connected correlation
function between spatial Wilson loops, by adopting substantially the same method used in \cite{SN03}.
Also for the glueball 2-point function we made use of an error reduction procedure based
on the idea proposed in \cite{lw}, and already applied for the computation of
glueball masses \cite{meyer}. We extracted the glueball masses at $t/a=2,3$
for $1/(aT_c)=2,3$ and $t/a=3,4$ for  $1/(aT_c)=4,6$.
Fig. \ref{mass_scaling} shows our final results for $m_{0^{++}}/T_c$ 
as function of $(aT_c)^2$; here the errors are dominated by the uncertainties
on  $m_{0^{++}}$.
By averaging several
results in the literature
\cite{glueb:teper98,Vaccarino:1999ku,Morningstar:1999rf,Liu:2001gx} for $m_{0^{++}}r_{0}$ and then using the continuum
limit relation \cite{SN03}
\begin{equation}
T_c r_0= 0.7498(50),
\end{equation}
we obtain
\begin{equation}\label{mct0cont}
m_{0^{++}}/T_c|_{a=0} =5.73(9) \;\;
\end{equation}
as estimation of the continuum result.
At $a\simeq 0.11 {\rm fm}$
we do  observe a moderate reduction of the lattice
artefacts by using $\beta_a<0$ with respect to the usual Wilson action
($\beta_a=0$). For $\beta_a=0$, the deviation from the continuum result of
eq.~(\ref{mct0cont}) amounts to $\sim 18\%$,
while for $\beta_a=-2,-4$ it
slightly decreases to  $\sim 12\%$.\\
At $a\simeq 0.17 {\rm fm}$ one observes
discretisation errors of $\sim 40\%$ for the Wilson action, while for the
mixed action they amount to  $\sim 25\%$ for $\beta_a=-2$ and $\sim 20\%$ for
$\beta_a=-4$.
\vspace{-0.2cm}
\section{CONCLUSIONS}
We investigated scaling properties of a SU(3) lattice gauge action with
plaquette terms in the fundamental and in the adjoint representation, with
negative adjoint coupling $\beta_a$. 
By studying the scaling behaviour of the quantity $T_c/\sqrt{\sigma}$ for the
different $\beta_a$ at our disposal, we did not observe a significant
improvement at negative adjoint couplings in comparison to the Wilson case
$\beta_a=0$. The values obtained with negative $\beta_a$ are a little
closer to the continuum limit. 
As expected, the mass $m_{0^{++}}$ of the lightest glueball is more 
sensitive to the variation of $\beta_a$. We investigated the scaling behaviour of the dimensionless quantity
$m_{0^{++}}/T_c$. Here indeed, we observed a significant reduction of the
lattice artefacts for negative $\beta_a$. At $a\simeq 0.17\, {\rm fm}$,
the lattice artefacts for $\beta_a=0$ are $~40\%$, while for $\beta_a=-4$ they
decrease to $~20\%$.


\end{document}